\documentclass{article}
\usepackage{spconf,amsmath,graphicx, amsfonts,url,cite}
\usepackage[normalem]{ulem}
\usepackage{color}
\usepackage{algorithm}
\usepackage{algorithmic}

\def\x{{\mathbf x}}
\def\h{{\mathbf h}}
\def\y{{\mathbf y}}
\def\Q{{\mathbf Q}}
\def\K{{\mathbf K}}
\def\V{{\mathbf V}}
\def\W{{\mathbf W}}
\def\R{{\mathbb R}}
\def\Z{{\mathbf Z}}

\def\head{\mathrm{head}}
\def\model{\mathrm{model}}
\def\Lblock{L_{\mathrm{block}}}
\def\Lhop{L_{\mathrm{hop}}}

\def\Energy{\mathrm{Energy}}
\def\ChunkEnergy{\mathrm{ChunkEnergy}}

\title{TOWARDS ONLINE END-TO-END TRANSFORMER AUTOMATIC SPEECH RECOGNITION}
\name{Emiru Tsunoo$^{1}$, Yosuke Kashiwagi$^{1}$, Toshiyuki Kumakura$^{1}$, Shinji Watanabe$^{2}$}
\address{$^1$Sony Corporation, Japan \\
$^2$Johns Hopkins University, USA}
\begin{document}
\ninept

\maketitle
\begin{abstract}
The Transformer self-attention network has recently shown promising performance as an alternative to recurrent neural networks in end-to-end (E2E) automatic speech recognition (ASR) systems.
However, Transformer has a drawback in that the entire input sequence is required to compute self-attention.
We have proposed a block processing method for the Transformer encoder by introducing a context-aware inheritance mechanism.
An additional context embedding vector handed over from the previously processed block helps to encode not only local acoustic information but also global linguistic, channel, and speaker attributes.
In this paper, we extend it towards an entire online E2E ASR system by introducing an online decoding process inspired by monotonic chunkwise attention (MoChA) into the Transformer decoder.
Our novel MoChA training and inference algorithms exploit the unique properties of Transformer, whose attentions are not always monotonic or peaky, and have multiple heads and residual connections of the decoder layers.
Evaluations of the Wall Street Journal (WSJ) and AISHELL-1 show that our proposed online Transformer decoder outperforms conventional chunkwise approaches.
\end{abstract}
\begin{keywords}
Speech Recognition, End-to-end, Transformer, Self-attention Network, Monotonic Chunkwise Attention
\end{keywords}
\section{Introduction}
\label{sec:intro}
End-to-end (E2E) automatic speech recognition (ASR) has been attracting attention as a method of directly integrating acoustic models (AMs) and language models (LMs) because of the simple training and efficient decoding procedures. 
In recent years, various models have been studied, including connectionist temporal classification (CTC) \cite{graves06, graves14, miao15, amodei16}, attention-based encoder--decoder models \cite{chorowski15, chan16, lu16, zeyer2018improved, chiu18}, their hybrid models \cite{kim17, watanabe17}, and the RNN-transducer \cite{graves12,graves13rnnt,rao17}.
Transformer \cite{vaswani17} has been successfully introduced into E2E ASR by replacing RNNs \cite{dong18, sperber18, salazar19, dong19, zhao19}, and it outperforms bidirectional RNN models in most tasks \cite{karita19}.
Transformer has multihead self-attention network (SAN) layers, which can leverage a combination of information from completely different positions of the input.

However, similarly to bidirectional RNN models \cite{schuster97}, Transformer has a drawback in that the entire utterance is required to compute self-attention, making it difficult to utilize in online recognition systems.
Also, the memory and computational requirements of Transformer grow quadratically with the input sequence length, which makes it difficult to apply to longer speech utterances.
A simple solution to these problems is block processing as in \cite{sperber18, dong19, jaitly2015neural}.
However, it loses global context information and its performance is degraded in general.

We have proposed a block processing method for the encoder--decoder Transformer model by introducing a context-aware inheritance mechanism, where an additional context embedding vector handed over from the previously processed block helps to encode not only local acoustic information but also global linguistic, channel, and speaker attributes \cite{tsunoo19}.
Although it outperforms naive blockwise encoders, the block processing method can only be applied to the encoder because it is difficult to apply to the decoder without knowing the optimal chunk step, which depends on the token unit granularity and the language.

For the attention decoder, various online processes have been proposed.
In \cite{chorowski15, chan16online, merboldt19}, the chunk window is shifted from an input position determined by the median or maximum of the attention distribution.  
Monotonic chunkwise attention (MoChA) uses a trainable monotonic energy function to shift the chunk window \cite{chiu2017monotonic}.
MoChA has also been extended to make it stable while training \cite{miao19} and to be able to change the chunk size adaptively to the circumstances \cite{fan19}.
\cite{moritz19} proposed a unique approach that uses a trigger mechanism to notify the timing of the attention computation.
However, to the best of our knowledge, such monotonic chunkwise approaches have not yet been applied to Transformer.

In this paper, we extend our previous context block approach towards an entire online E2E ASR system by introducing an online decoding process inspired by MoChA into the Transformer decoder.
Our contributions are as follows.
1) Triggers for shifting chunks are estimated from the source--target attention (STA), which uses queries and keys,
2) all the past information is utilized according to the characteristics of the Transformer attentions that are not always monotonic or locally peaky, and
3) a novel training algorithm of MoChA is proposed, which extends to train the trigger function by dealing with multiple attention heads and residual connections of the decoder layers.
Evaluations of the Wall Street Journal (WSJ) and AISHELL-1 show that our proposed online Transformer decoder outperforms conventional chunkwise approaches.

\section{Transformer ASR}
\label{sec:transformer}
The baseline Transformer ASR follows that in \cite{karita19}, which is based on the encoder--decoder architecture. 
An encoder transforms a $T$-length speech feature sequence $\x = (x_{1},\dots,x_{T})$ to an $L$-length intermediate representation $\h = (h_{1},\dots,h_{L})$, where $L \leq T$ due to downsampling.
Given $\h$ and previously emitted character outputs $\y_{i-1} = (y_{1},\dots,y_{i-1})$, a decoder estimates the next character $y_{i}$.


The encoder consists of two convolutional layers with stride $2$ for downsampling, a linear projection layer, positional encoding, followed by $N_{e}$ encoder layers and layer normalization.
Each encoder layer has a multihead SAN followed by a position-wise feedforward network, both of which have residual connections.
Layer normalization is also applied before each module.
In the SAN, attention weights are formed from queries ($\mathbf{Q} \in \R^{t_q\times d}$) and keys ($\mathbf{K} \in \R^{t_k\times d}$), and applied to values ($\mathbf{V} \in \R^{t_v\times d}$) as
\vspace{-1mm}
\begin{align}
    \mathrm{Attention}(\Q,\K,\V) = \mathrm{softmax}\left(\frac{\Q\K^T}{\sqrt{d}}\right)\V, \label{eq:attention}
\end{align}
where typically $d = d_{\model}/M$ for the number of heads $M$.
We utilized multihead attention denoted, as the $\mathrm{MHD}(\cdot)$ function, as follows:
\vspace{-1mm}
\begin{align}
    & \mathrm{MHD}(\Q,\K,\V) = \mathrm{Concat}(\head_{1},\dots,\head_{M})\W_O^n,
    \label{eq:mhead} \\
    & \head_{m} = \mathrm{Attention}(\Q\W_{Q,m}^n,\K\W_{K,m}^n,\V\W_{V,m}^n). \label{eq:head}
\end{align}

In \eqref{eq:mhead} and \eqref{eq:head}, the $n$th layer is computed with the projection matrices $\W_{Q,m}^n \in \R^{d_{\model} \times d}$, $\W_{K,m}^n \in \R^{d_{\model} \times d}$, $\W_{V,m}^n \in \R^{d_{\model} \times d}$, and $\W_{O}^n \in \R^{Md \times d_{\model}}$.
For all the SANs in the encoder, $\Q$, $\K$, and $\V$ are the same matrices, which are the inputs of the SAN. 
The position-wise feedforward network is a stack of linear layers.

The decoder predicts the probability of the following character from previous output characters $\y_{i-1}$ and the encoder output $\h$, i.e., $p(y_i|\y_{i-1},\h)$.
The character history sequence is converted to character embeddings.
Then, $N_{d}$ decoder layers are applied, followed by the linear projection and Softmax function.
The decoder layer consists of a SAN and a STA, followed by a position-wise feedforward network.
The first SAN in each decoder layer applies attention weights to the input character sequence, where the input sequence of the SAN is set as $\Q$, $\K$, and $\V$.
Then, the following STA attends to the entire encoder output sequence by setting $\K$ and $\V$ to be the encoder output $\h$.

The SAN can leverage a combination of information from completely different positions of the input.
This is due to the multiple heads and residual connections of the layers that complement each other, i.e., some attend {\it monotonically and locally} while others attend {\it globally}.
Transformer requires the entire speech utterance for both the encoder and decoder; thus, they are processed only after the end of the utterance, which causes a huge delay.
To realize an online ASR system, both the encoder and decoder are processed online.


\begin{figure}[t]
  \centering
  \includegraphics[width=0.9\columnwidth]{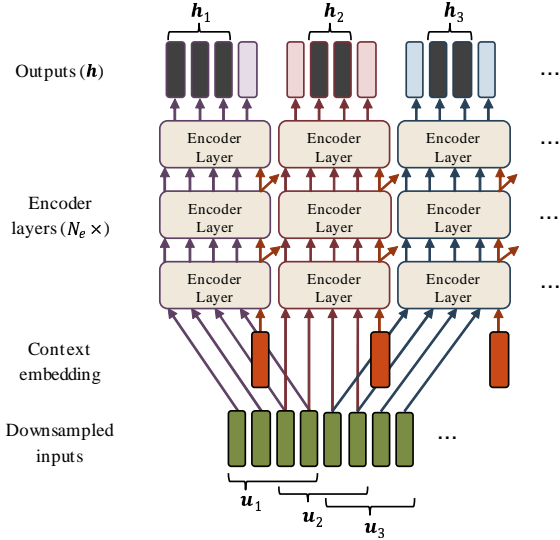}
  \vspace{-0.5cm}
  \caption{Context inheritance mechanism of the encoder.}
  \label{fig:context}
\end{figure}

\section{Contextual Block Processing of Encoder}
\label{sec:encoder}
A simple way to process the encoder online is blockwise computation, as in \cite{sperber18, dong19, jaitly2015neural}.
However, the global channel, speaker, and linguistic context are also important for local phoneme classification.
We have proposed a context inheritance mechanism for block processing by introducing an additional context embedding vector\cite{tsunoo19}.
As shown in the tilted arrows in Fig.~\ref{fig:context}, the context embedding vector is computed in each layer of each block and handed over to the upper layer of the following block.
Thus, the SAN in each layer is applied to the block input sequence using the context embedding vector.

The context embedding vector is introduced into the original formulation in Sec.~\ref{sec:transformer}.
Denoting the context embedding vector as $\mathbf{c}_{b}^{n}$, the augmented variables satisfy $\Tilde{\Q}_b^{n} = [\Z_b^{n-1} \ c_{b}^{n-1}]$ and $\Tilde{\K}_b^{n}=\Tilde{\V}_b^n=[\Z_b^{n-1} \ c_{b-1}^{n-1}]$, where the context embedding vector of the previous block $(b-1)$ of the previous layer $(n-1)$ is used.
$\Z_b^{n}$ is the output of the $n$th encoder layer of block $b$, which is computed simultaneously with the context embedding vector $c_b^{n}$ as 
\vspace{-1mm}
\begin{align}
    [\Z_b^{n} \ c_b^{n}] &= \max(0, \Tilde{\Z}_{b,\text{int.}}^{n}\W_1^n + v_1^n)\W_2^n + v_2^n + \Tilde{\Z}_{b,\text{int.}}^{n} \\
    \Tilde{\Z}_{b,\text{int.}}^{n} &= \mathrm{MHD}(\Tilde{\Q}_b^{n},\Tilde{\K}_b^{n},\Tilde{\V}_b^{n}) + \Tilde{\V}_{b}^{n},
\end{align}
where $\W_1^n$, $\W_2^n$, $v_1^n$, and $v_2^n$ are trainable matrices and biases.
The output of the SAN does not only encode input acoustic features but also delivers the context information to the succeeding layer as shown by the tilted red arrows in Fig.~\ref{fig:context}.

\section{Online Process for Decoder}
\label{sec:decoder}
\subsection{Online Transformer Decoder based on MoChA}
\label{ssec:mocha}
\begin{figure}[t]
  \centering
  \includegraphics[width=1\columnwidth]{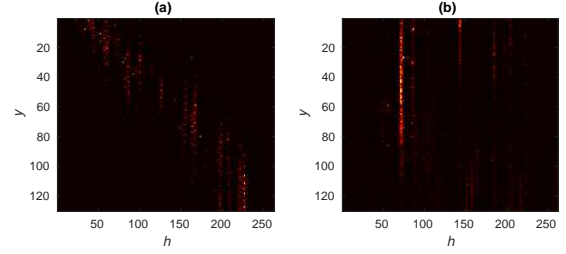}
  \vspace{-0.5cm}
  \caption{Examples of attentions in a Transformer decoder layer. (a) is a head having wider attentions, and (b) is a head attending a certain area of $\h$.}
  \label{fig:wide}
\end{figure}

The decoder of Transformer ASR is incremental at test time, especially for the first SAN of each decoder layer.
However, the second STA requires the entire sequence of the encoded features $\h$.
Blockwise attention mechanisms cannot be simply applied with a fixed step size, because the step size depends on the output token granularity (grapheme, character, (sub-)word, and so forth) and language.
In addition, not all the STAs are monotonic, because the other heads and layers complement each other.
Typically, in the lower layer of the Transformer decoder, some heads attend wider areas, and some attend a certain area constantly, as shown in Fig.~\ref{fig:wide}.
Therefore, chunk shifting and the chunk size should be adaptive.

For RNN models, the median or maximum of the attention distribution is used as a cue for shifting a fixed-length chunk, where the parameters of the original batch models are reused \cite{chorowski15, chan16online, merboldt19}.
MoChA further introduces the probability distribution of chunking to train the monotonic chunking mechanism.
In this paper, we propose a novel online decoding method inspired by MoChA.

MoChA \cite{chiu2017monotonic} splits the input sequence into small chunks over which soft attention is computed.
It learns a monotonic alignment between the encoder features $\h$ and the output sequence $\y$, with $w$-length chunking.
``Soft'' attention is efficiently utilized with backpropagation to train chunking parameters.
At the test time, online ``hard'' chunking is used to realize online ASR, which achieves almost the same performance as the soft attention model.


\begin{algorithm}[t]                      
\caption{MoChA Inference for $n$-th Transformer Decoder Layer}         
\label{alg:inference}                          
\begin{algorithmic}[1]     
\REQUIRE encoder features $\h$, length $L$, chunk size $w$
\STATE \textbf{Initialize:} $y_0=\langle sos\rangle$, $t_{m,0}=1$, $i=1$
\WHILE{$y_{i-1} \neq \langle eos \rangle$}
\FOR{\textcolor{red}{$m=1$ \TO $M$}}
\FOR{$j=t_{m,i-1}$ \TO $L$}
\STATE $p_{m,i,j}=\sigma(\textcolor{red}{\Energy_m}(z_{\mathrm{SAN},i},h_{j}))$
\IF{$p_{m,i,j} \geq 0.5$}
\STATE $t_{m,i}=j$
\STATE \textbf{break}
\ENDIF
\ENDFOR
\IF{$p_{m,i,j}<0.5, \forall j \in \{t_{m,i-1},\dots,L\}$}
\STATE \textcolor{red}{$t_{m,i}=t_{m,i-1}$}
\ENDIF
\STATE $r=t_{m,i}-w+1$ \ \ \textcolor{red}{// or $r=1$}
\FOR{$k=r$ \TO $t_{i}$}
\STATE{$u_{m,i,k}=\textcolor{red}{\ChunkEnergy_m}(z_{\mathrm{SAN},i},h_{k})$}
\ENDFOR
\STATE $\head_{m,i}=\sum_{k=r}^{t_{i}}\frac{\exp(u_{i,k})}{\sum_{l=r}^{t_{i}}\exp(u_{i,l})}v_{m,k}$
\ENDFOR
\STATE $z_{\mathrm{STA},i}=\textcolor{red}{\mathrm{STA}(\y_{i-1},\head_{1,i},\dots,\head_{M,i})}$, $i=i+1$
\ENDWHILE
\end{algorithmic}
\end{algorithm}

Since Transformer has unique properties, the conventional MoChA cannot be simply applied.
One property is that the STA is computed using queries and keys, while MoChA is formulated on the basis of the attention using a hidden vector of the RNN and $\tanh$.
Another property is that not all the STAs are monotonic, because the other heads and layers complement each other, as examples shown in Fig.~\ref{fig:wide}.
We modify the training algorithm of MoChA to deal with these characteristics.

\subsection{Inference Algorithm}
\label{ssec:inference}
The inference process for decoder layer $n$ is shown in Algorithm~\ref{alg:inference}.
The differences from the original MoChA are highlighted in red color.
In our case, MoChA decoding is introduced into the second STA of each decoder layer; the vector $z_{\mathrm{SAN},i}$ in Algorithm~\ref{alg:inference} is the output of the first SAN in the decoder layer.
$\mathrm{STA}(\cdot)$ in line 20 concatenates and computes an output of the STA network, $z_{\mathrm{STA},i}$, in each decoder layer, as in (\ref{eq:mhead}).
MoChA can be applied independently to each head; thus, we added line 3.
In line 18, the attention weight is applied to the selected values $v_{m,k} = h_{k}W_{V,m}$ to compute $\head_m$ in (\ref{eq:head}), and the chunk of selection shifts monotonically.

$p_{m,i,j}$ in line 5 is regarded as a trigger function at head $m$ to move the computing chunk, which is estimated from an $\Energy$ function.
For the $\Energy$ and $\ChunkEnergy$ (in line 16) functions, the original MoChA utilizes $\tanh$ because it is used as a nonlinear function in RNNs.
However, in Transformer, attentions are computed using queries and keys as in (\ref{eq:attention}).
Therefore, we modify them for the head $m$ as 
\vspace{-1mm}
\begin{align}
    \Energy_m(z_{\mathrm{SAN},i},h_{j}) &= g_{m}\frac{q_{i,m}k_{j,m}^T}{\sqrt{d}||q_{i,m}||} + r_{m},
\end{align}
\begin{align}
    \ChunkEnergy_m(z_{\mathrm{SAN},i},h_{j}) &= \frac{q_{i,m}k_{j,m}^T}{\sqrt{d}},
\end{align}
where $g_{m}$ and $r_{m}$ are trainable scalar parameters, $q_{i,m}=z_{\mathrm{SAN},i}W_{Q,m}$, and $k_{j,m}=h_{j}W_{K,m}$ as in (\ref{eq:head}).

Note that, the exception in lines 11--13, where the trigger never ignites in frame $i$, sets $\head_{m,i}$ as $\mathbf{0}$ in the original MoChA.
However, we compute $\head_{m,i}$ using the previous $t_{m,i-1}$ (line 12) because the exception often occurs in Transformer.
Also, for online processing, all the past frames of encoded features $\h$ are also available without any latency, while the original MoChA computes attentions within the fixed-length chunk.
Taking into account the property that Transformer attentions tend to be distributed widely and are not always monotonic, we also consider utilizing the past frames.
We optionally modify line 14 by setting $r=1$ and test both cases in Sec.~\ref{sec:experiments}.

\subsection{Training Algorithm}
\label{ssec:training}
MoChA strongly relies on the monotonicity of the attentions, and it also forces attentions to be monotonic, while Transformer has a flexible attention mechanism that may be able to integrate information of various positions without the monotonicity.
Further more, the Transformer decoder has both multihead and residual connections.
Therefore, typically, not all the attentions become monotonic, as in Fig.~\ref{fig:wide}.

\begin{algorithm}[t]                      
\caption{MoChA Training for $n$-th Transformer Decoder Layer}   
\label{alg:train}                          
\begin{algorithmic}[1]     
\REQUIRE encoder features $\h$, length $L$, chunk size $w$, Gauss. noise $\epsilon$
\STATE \textbf{Initialize:} $y_0=\langle sos\rangle$, $\alpha_{0,0}=1$, $\alpha_{0,k}=0 (k\neq 0)$, $i=1$
\WHILE{$y_{i-1} \neq \langle eos \rangle$}
\FOR{\textcolor{red}{$m=1$ \TO $M$}}
\FOR{$j=1$ \TO $L$}
\STATE $p_{m,i,j}=\sigma(\textcolor{red}{\Energy_m}(z_{\mathrm{SAN},i},h_{j}) + \epsilon)$
\STATE \textcolor{red}{$q_{m,i,j}=\prod_{k=j+1}^{L}(1-p_{m,i,k})$}
\STATE \textcolor{red}{$\alpha_{m,i,j}=p_{m,i,j}\sum_{k=1}^{j}\left(\alpha_{m,i-1,k}\prod_{l=k}^{j-1}(1-p_{m,i,l})\right)$} \\ \ \ \ \ \ \ \ \ \ \ \ \ \ \ \ \ \ \ \ \ $\textcolor{red}{+ q_{m,i,j}\alpha_{m,i-1,j}}$
\ENDFOR
\FOR{$j=1$ \TO $L$}
\STATE $u_{m,i,k}=\textcolor{red}{\ChunkEnergy_m}(z_{\mathrm{SAN},i},h_{k})$
\STATE $\beta_{m,i,j}=\sum_{k=j}^{j+w-1}\frac{\alpha_{m,i,j}\exp(u_{m,i,k})}{\sum_{l=k-w+1}^{k}\exp(u_{m,i,l})}$
\ENDFOR
\STATE $\head_{m,i}=\sum_{j=1}^{L}\beta_{m,i,j}v_{j}$
\ENDFOR
\STATE $z_{\mathrm{STA},i}=\textcolor{red}{\mathrm{STA}(y_{i-1},\head_{1,i},\dots,\head_{M,i})}$, $i=i+1$
\ENDWHILE
\end{algorithmic}
\end{algorithm}

\begin{figure}[t]
  \hspace{-0.7cm}
  \includegraphics[width=1.1\columnwidth]{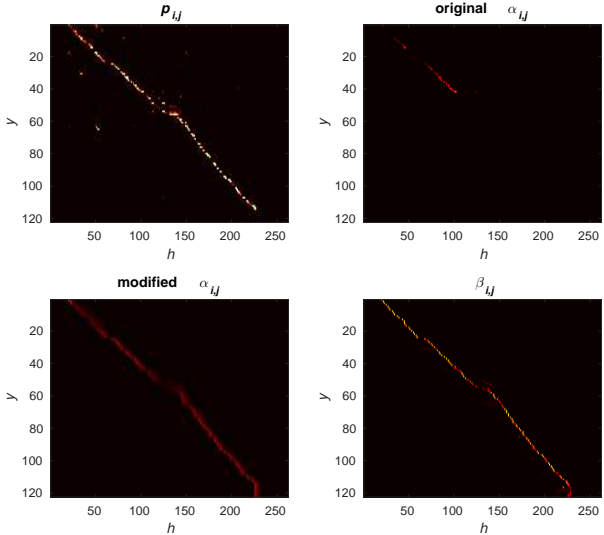}
  \vspace{-1.0cm}
  \caption{Example of expected attention in the Transformer decoder. Top left: $p_{i,j}$ in Algorithm~\ref{alg:train}; top right: original $\alpha_{i,j}$ in (\ref{eq:orgalpha}); bottom left: our modified $\alpha_{i,j}$ in Algorithm~\ref{alg:train}; bottom right: expected attention $\beta_{i,j}$. Head index $m$ is omitted for simplicity.}
  \label{fig:attention}
\end{figure}

The original MoChA training computes a variable $\alpha_{i,j}$, which is a cumulative probability of computing the local chunk attention at $t_{i}=j$, defined as
\vspace{-1mm}
\begin{align}
    \alpha_{i,j} = p_{i,j}\sum_{k=1}^j\left(\alpha_{i-1,k}\prod_{l=k}^{j-1}(1-p_{i,l})\right). \label{eq:orgalpha}
\end{align}
When $p_{i,j}\approx0$ for all $j$, which occurs frequently in Transformer because the other heads and layers complement each other for this frame, $\alpha_{i,j}$ rapidly decays after $i$.
An example is shown in Fig.~\ref{fig:attention}.
The top left shows $p_{m,i,j}$ in Algorithm~\ref{alg:inference}, which has monotonicity. 
The top right is the original $\alpha_{i,j}$ in (\ref{eq:orgalpha}), in which the value decreases immediately after around frame 50 of the target $y$ and does not recover.

Therefore, we introduce a probability of the trigger not igniting as $q_{m,i,j}$ into computation of $\alpha_{m,i,j}$.
Thus, the new training algorithm for Transformer is shown in Algorithm~\ref{alg:train}, which encourages MoChA to exploit the flexibility of the SAN in Transformer (colored lines are new to the original MoChA).
An example of our modified $\alpha_{m,i,j}$ is shown in the bottom left of Fig.~\ref{fig:attention}, which maintains the monotonicity. 
The bottom right shows the expected attention $\beta_{m,i,j}$.

\section{Experiments}
\label{sec:experiments}

\subsection{Experimental Setup}
\label{ssec:setups}
We carried out experiments using the WSJ English and AISHELL-1 Mandarin data \cite{aishell17}.
The input acoustic features were 80-dimensional filter banks and the pitch, extracted with a hop size of 10 ms and a window size of 25 ms, which were normalized with the global mean and variance.
For the WSJ English setup, the number of output classes was 52, including symbols.
We used 4,231 character classes for the AISHELL-1 Mandarin setup.

For the training, we utilized multitask learning with CTC loss as in \cite{watanabe17,karita19} with a weight of 0.1.
A linear layer was added onto the encoder to project $\h$ to the character probability for the CTC. 
The Transformer models were trained over 100 epochs for WSJ and 50 epochs for AISHELL-1, with the Adam optimizer and Noam learning rate decay as in \cite{vaswani17}.
The learning rate was set to 5.0 and the minibatch size to 20.
SpecAugment \cite{park19} was applied to only WSJ.

The parameters of the last 10 epochs were averaged and used for inference.
The encoder had $N_{e}=12$ layers with 2048 units and the decoder had $N_{d}=6$ layers with 2048 units, with both having a dropout rate of 0.1.
We set $d_{model}=256$ and $M=4$ for the multihead attentions.
We trained three types of Transformer, namely, baseline Transformer \cite{karita19}, Transformer with the contextual block processing encoder (CBP Enc.\ + Batch Dec.) \cite{tsunoo19}, and the proposed entire online model with the online decoder (CBP Enc.\ + Proposed Dec.).
The training was carried out using ESPNet \cite{watanabeespnet} with the PyTorch backend.
The median based chunk shifting \cite{chorowski15} with a window of 16 frames was also applied to the Batch Dec.\ with and without past frames for the fair comparison (CBP Enc.\ + Median Dec.). 

For the CBP Enc.\ models, we set the parameters as $\Lblock=16$ and $\Lhop=8$.
For the initialization of context embedding, we utilized the average of the input features to simplify the implementation.
The decoder was trained with the proposed MoChA architecture using $w=8$.
The STA were computed within each chunk, or using all the past frames of encoded features as described in Sec. \ref{ssec:inference}.

The decoding was performed alongside the CTC, whose probabilities were added with weights of 0.3 for WSJ and 0.7 for AISHELL-1 to those of Transformer.
We performed decoding using a beam search with a beam size of 10.
An external word-level LM, which was a single-layer LSTM with 1000 units, was used for rescoring using shallow fusion \cite{kannan18} with a weight of $1.0$ for WSJ.
A character-level LM with the same structure was fused with a weight of $0.5$ for AISHELL-1.

For comparison, unidirectional and bidirectional LSTM models were also trained as in \cite{watanabe17}.
The models consisted of an encoder with a VGG layer, followed by LSTM layers and a decoder.
The numbers of encoder layers were six and three, with 320 and 1024 units for WSJ and AISHELL-1, respectively.
The decoders were an LSTM layer with 300 units for WSJ and two LSTM layers with 1024 units for AISHELL-1.

\begin{table}[t]
  \caption{Word error rates (WERs) in the WSJ and AISHELL-1 evaluation task.}
  \label{tab:result}
  \vspace{1mm}
  \centering
  \scalebox{0.9}{
  \begin{tabular}{l|cc}
    \hline
     & WSJ (WER) & AISHELL-1 (CER) \\
    \hline\hline
    \multicolumn{2}{l}{Batch processing}  \\
    \hline
    biLSTM \cite{watanabe17} & 6.7 & 9.2 \\
    uniLSTM & 8.4 & 11.8 \\
    Transformer \cite{karita19} & 4.9 & 6.7 \\
    CBP Enc. + Batch Dec. \cite{tsunoo19} & 6.0 & 7.6\\
    \hline
    \multicolumn{2}{l}{Online processing} \\
    \hline
    CBP Enc. + median Dec. \cite{chorowski15} & 9.9 & 25.0\\
     \ \ ---{\it with past frames} & 7.9 & 24.2 \\
    CBP Enc. + Proposed Dec. & 8.8 & 18.7\\
    \ \ ---{\it with past frames} & {\bf 6.6} & {\bf 9.7} \\
    \hline
  \end{tabular}
  }
\end{table}

\subsection{Results}
\label{ssec:results}
Experimental results are summarized in Table~\ref{tab:result}.
The chunk hopping using the median of attention worked well in the English task but poorly in the Chinese task.
This was because Chinese requires a wider area of the encoded features to emit each character.
On the other hand, our proposed decoder prevented the degradation of performance. 
In particular, using all the past frames of encoded features, our proposed decoder achieved the highest accuracy among the online processing methods. 
This indicated that the new decoding algorithm was able to exploit the wider attentions of Transformer.

\section{Conclusion}
\label{sec:conclusion}
We extended our previous Transformer, which adopted a contextual block processing encoder, towards an entirely online E2E ASR system by introducing an online decoding process inspired by MoChA into the Transformer decoder.
The MoChA training and inference algorithms were extended to cope with the unique properties of Transformer whose attentions are not always monotonic or peaky and have multiple heads and residual connections of the decoder layers.
Evaluations of WSJ and AISHELL-1 showed that our proposed online Transformer decoder outperformed conventional chunkwise approaches.
Thus, we realize the entire online processing of Transformer ASR with reasonable performance.

\bibliographystyle{IEEEbib}
\bibliography{mybib}

\end{document}